\documentclass[aps,prb,twocolumn,showpacs,letterpaper]{revtex4}
\usepackage{graphicx}
\usepackage{subfigure}
\usepackage{amsmath}
\usepackage{color}

\newcommand{\aF}{$\alpha^2 F(\omega)$}
\newcommand{\aFIJ}{$\alpha^2F_{JJ'} (\omega)$}

\newcommand{\be}{\begin{equation}}
\newcommand{\ee}{\end{equation}}
\newcommand{\bea}{\begin{eqnarray}}
\newcommand{\eea}{\end{eqnarray}}
\newcommand{\ba}{\begin{array}}
\newcommand{\ea}{\end{array}}

\newcommand{\cac}{{CaC$_6$}}

\begin{document}

%================================
% TITLE
%================================
\title{Phononic self-energy effects and superconductivity in CaC$_{\bf 6}$ } 

%================================
% LIST OF AUTHORS
%================================

\author{A. Sanna$^{1,2}$, S. Pittalis$^3$, J. K. Dewhurst$^{1,2}$, M. Monni$^4$,\\ S. Sharma$^{1,2}$, G. Ummarino$^5$, S. Massidda$^4$ and E. K. U. Gross$^{1,2}$}
\affiliation{$^1$ Max-Planck-Institut f\"ur Mikrostrukturphysik, Weinberg 2, D-06120 Halle (Germany)}
\affiliation{$^2$ European Theoretical Spectroscopy Facility (ETSF)}
\affiliation{$^3$Department of Physics and Astronomy,  University of Missouri--Columbia, Columbia MO, 65211 (USA)}
\affiliation{$^4$Dipartimento di Scienze Fisiche, Universit\`a degli Studi di Cagliari, Cagliari (Italy)}
\affiliation{$^5$C.N.I.S.M. â Dipartimento di Fisicaâ Politecnico di Torino,  C.so Duca degli Abruzzi 24, 10129 Torino  (Italy)}

\begin{abstract}

We study the graphite intercalated compound \cac\, by means of Eliashberg theory. We perform an analysis of the electron-phonon coupling and define a minimal $6$-band anisotropic structure, that leads to a Fermi surface dependance of the superconducting gap.  
A comparison of the superconducting gap structure obtained using the Eliashberg and the superconducting density functional theory is performed.
We further report the anisotropic properties of the electronic spectral function, the polaronic quasi-particle branches and their interplay with Bogoljubov excitations. 
 
\noindent

\end{abstract}


\maketitle
\newpage
%================================
%\section{Introduction}
%\label{sec1}
%================================

The electron-phonon interaction leads to many significant physical phenomenon in solids (notably, superconductivity), and has therefore been studied extensively both in model systems and in real materials. One important aspect of this kind of interaction is the formation of a coupled electron-phonon system with new interesting features such as the appearance of polaronic sub-bands branching from the main electronic bands. This low  energy features of the electronic structure can be observed thanks to the recent developments in the resolution of Angle-Resolved Photo Emission Spectroscopy (ARPES).

%Recent developments of Angle-Resolved Photo Emission Spectroscopy (ARPES) allows for the investigation of such low energy contributions, to the electronic structure, from the electron-phonon interaction.

The theoretical background to deal with metallic polarons has been laid down by Engelsberg and Schrieffer (ES)~\cite{EngelsbergSchrieffer}. For this they used a field theoretical approach combined with Einstein-Debye model to mimic the phonon spectrum. 
The ES theory shows the damping of electrons by phonons and the development of branches in the electronic dispersion corresponding to energy and strength of the phonon modes.
In superconductors, the electron-phonon interaction leads to the formation of a superconducting gap below the critical temperature $T_c$. 
These phenomena can be well described within the Eliashberg theory which extends the ES theory to the superconducting state~\cite{Eliashberg,ScalapinoSchriefferWilkins} and reduces to the ES theory in the non-superconducting normal state.

Due to several computational complexities, a proper account of the material specific electronic and phononic structures could not be achieved until very recently-- Eiguren {\em et al.}~\cite{Eiguren-1} and Eiguren and Ambrosch-Draxl~\cite{eiguren-2} studied the effect of the electron-phonon interaction on the electronic self-energy in the normal state.
The main properties of the spectral function in the superconducting state have been reported by Scalapino~\cite{ScalapinoSchriefferWilkins,ScalapinoParksSC} and, more recently have also been studied using the ARPES experimental data~\cite{Sandvik,Devereaux,Zhou_HTSC}.
However, to the best of our knowledge, no first-principles attempt has been made to study the effect of polarons in the superconducting state.

In the present work we use the Eliashberg~\cite{Eliashberg,AllenMitrovic,Schrieffer,ScalapinoSchriefferWilkins} method to study the behavior of ES polarons; a detailed  analysis of the electronic self-energy, including electron-phonon contributions is performed.
In particular, the features originating from the anisotropy of the electron-phonon coupling are investigated.
Most importantly, it is shown how the polaronic branches change in the superconducting state below $T_{c}$.

The system that is considered for this analysis is the graphite intercalated compound \cac\, 
This material has the highest superconducting  $T_c$ observed so far ($11.5$ K) amongst the group of graphitic compounds.
Graphite related materialas have attracted considerable interest in the last few years, mostly due to the appealing possibility of tuning their physical properties~\cite{Dresselhaus}. 
In particular it is possible to vary the conductivity of graphite from semi-metallic\cite{Review_Graphite} to metallic and to superconducting~\cite{Hannay,Weller,Emery,Csanyi,CalandraMauri,Mazin_CaC6_YbC6,Boeri_GICelph,Kim_GIC,MazinBoeri_problems} by adjusting the level of intercalation.
In Ca intercalated graphite superconductivity arises from the strong electron-phonon coupling provided both by C and Ca phonon modes~\cite{CalandraMauri,CalandraMauri2}. This coupling is strongly anisotropic with C and Ca related  phonons acting selectively on the multiple Fermi Surface (FS) sheets of the system~\cite{CaC6-nostro}. 
%These peculiarities of the electron-phonon interaction reflect on the electron phonon coupling, making this system particularly interesting.
These peculiarities make the system particularly interesting.

The paper is organized as follows: In Sec.~I,  the main concepts and physical quantities describing our results are introduced by reviewing the Eliashberg theory of superconductivity. 
Sec.~II is devoted to a detailed description of the principal computational techniques employed in this work. 
In Sec.~III, results for \cac\ are discussed.  Sec.~IIIA reports on the structure of the electron-phonon interaction. Sec.~IIIB focuses on the numerical solutions of the Eliashberg equations with a $\mu^*$ determined using results from Density Functional Theory for Superconductors (SCDFT) calculations.
In Sec.~IIIC, the polaronic features of the excitation spectrum of \cac\ are elucidated, both in the normal state (Sec.~IIIC1) and in the superconducting state (Sec.~IIIC2). Finally, conclusions are drawn in Sec.~IV.

%========================
\section {METHODS}
\label{sec2}
%========================
Central quantity in the Nambu-Gor'kov formalism of superconductivity is the $2\times2$ Green's function~\cite{Schrieffer}:
\begin{equation}\label{eq:symGfermion}
\bar G({\bf k},i\omega_n) \equiv \left(\begin{array}{ccc}
G({\bf k},i\omega_n)&-F({\bf k},i\omega_n)\\-F^*({\bf k},i\omega_n)&G({\bf k},-i\omega_n) \end{array} \right),
\end{equation}
where $G({\bf k},i\omega_n)$ and $F({\bf k},i\omega_n)$ are, respectively, the normal and anomalous electronic Green's functions in reciprocal space.
$\omega_n$ are the Fermionic Matsubara frequencies given by $\omega_n=\pi(2n+1)k_BT$ with $T$ being the temperature and $k_B$ the Boltzmann constant. 
Following a well established procedure\cite{AllenMitrovic}, non-interating Kohn-Sham system with Green's function $\bar G_0({\bf k},i\omega_n)=\left[i\omega_n \sigma_0 - \xi_{{\bf k}}\sigma_3\right]^{-1}$, is used as a starting point. Here $\sigma_j$~($j = 0~...~3$) are the Pauli matrices and $\xi_{{\bf k}}$ are the Kohn-Sham eigenvalues relative to the Fermi energy.
The interacting Green's function can then be obtained using perturbation theory: 
\begin{equation}\label{eq:Dyson}
\left[\bar G({\bf k},i\omega_n)\right]^{-1}=\left[\bar G_0({\bf k},i\omega_n)\right]^{-1}-\bar \Sigma({\bf k},i\omega_n)\;.
\end{equation} 
The following approximation for the electronic self-energy is used:
 \begin{eqnarray}\label{eq:sigma}
\bar\Sigma({\bf k},i\omega_n) &=& -k_BT\sum_{{\bf k}',n'}\sigma_3\bar G({\bf k}',i\omega_{n'})\sigma_3   \nonumber \\
&\times& \left[\sum_{\nu}\left|g_{{\bf k},{\bf k}',\nu}\right|^2D_{\nu}({\bf k}-{\bf k}',i\omega_n-i\omega_{n'})  \right. \nonumber \\
&+&  \left.  \sigma_1W({\bf k}-{\bf k}') \right]
\end{eqnarray}
here $D$ is the phonon propagator ($D={-2\omega_{{\bf q},\nu}}/{\left(\omega_n^2+\omega^2_{{\bf q},\nu}\right)}$), $g_{{\bf k},{\bf k}',\nu}$ are the electron-phonon matrix elements\cite{BaroniRMP} between states with wavevector ${\bf k}$ and ${\bf k}'$, and due to a phonon mode of index $\nu$ and wavevector ${\bf q}$=${\bf k}-{\bf k}'$ and $\omega_{{\bf q},\nu}$ is the frequency of the mode obtained via linear response~\cite{BaroniRMP} of the Kohn-Sham system. This way of calculating $\omega_{{\bf q},\nu}$ is known to lead to a very good agreement with the measured phononic branches, at least for standard metals and insulators~\cite{BaroniRMP}. 
$W$ in Eq. \ref{eq:sigma} is the screened static electron-electron interaction, it accounts for those parts of the interaction which do not involve any phononic contribution.  
The $\sigma_1$ factor in front of $W$ accounts for the fact that exchange and correlation effects are already included in $\bar G_0({\bf k},i\omega_n)$\cite{Marini_Cu}. Then only off-diagonal contributions of $W$ are retained in $\bar \Sigma$.

%%%%%%%%%%%%%%%%%%%%%%%%%%%%%%%%%%%%%%%%
The treatment of the Coulomb term needs particular care.
Within Eliashberg theory, an arbitrary cut-off in ${\bf k}$ space is needed in order to avoid serious convergency problems in the Matsubara summation~\cite{Morel_Anderson,ScalapinoSchriefferWilkins}.
A conventional way to deal with this problem is to choose an energy cut-off of the order of the Fermi energy, 
and to assume that the product of $W$ with the density of states (DOS) is constant: $\mu = WN$, $N$ being the 
DOS per spin at the Fermi energy.
It is then possible to restrict the Matsubara integration to a low energy ( a fraction of eV ) by a renormalization procedure $\mu \rightarrow \mu^*$ introduced by Morel and Anderson\cite{ScalapinoSchriefferWilkins,LeavensFenton,Morel_Anderson}. 
$\mu^*$ can be calculated within the random phase approximation~\cite{CaC6-nostro,mu_MgB2,CaC6-H} (RPA) and the resulting $T_c$ are usually in reasonable agreement with experiments~\cite{AllenMitrovic,Carbotte}. However, usually $\mu^*$ is adjusted to obtain the experimental $T_c$. 
%%%%%%%%%%%%%%%%%%%%%%%%%%%%%%%%%%%%%%%%

The self-energy in Eq.~(\ref{eq:sigma}) is ${\bf k}$-dependent and leads to anisotropic Eliashberg equations which are computationally very demanding. 
A simplification can be introduced retaining a minimal anisotropic structure needed for the properties of interest.
The FS may be divided into portions (FS sheets), with each sheet identifying a corresponding intersecting energy band. 
These FS sheets and energy bands can be labeled using the same index  (say, $J$). We shall refer to such a division
as to a {\em multi-band} decomposition (details of our procedure are given in Sec. \ref{sec:res-elph}).

The electron-phonon coupling can be averaged over a prescribed multi-band divisions  (see below, Eq.s~(\ref{el5}) and (\ref{eq:a2f})). 
Corresponding, multi-band resolved self-energy, $\Sigma_J$, expanded in the basis of Pauli matrices has a form:
\begin{equation}\label{eq:SE_expanded}
\Sigma_J(n)=i\omega_n\left[1-Z_J(n)\right] \sigma_0 + \Delta_J(n)Z_J(n)\sigma_1\;,
\end{equation}
(terms which only result in a rigid shift of the Fermi level are neglected).
The multi-band resolved Green's function reads:
\begin{equation}\label{eq:Gmatsubara}
\bar G_J({\bf k},i\omega_n)=\frac{-\left(\begin{array}{ccc}
i\omega_nZ_J(n)+\xi_{J{\bf k}}&\Delta_J(n)Z_J(n)\\
\Delta_J(n)Z_J(n)&i\omega_nZ_J(n)-\xi_{J{\bf k}}\end{array} \right)}{\left[Z_J(n)\omega_n\right]^2+\xi^2_{J{\bf k}}+\left[\Delta_J(n)Z_J(n)\right]^2},
\end{equation}
where $\xi_{J{\bf k}}$ are the Kohn-Sham eigenvalues in the $J$-th band.
Using Eqs.~(\ref{eq:SE_expanded}) and (\ref{eq:Gmatsubara}) in the Dyson equation, we arrive at the following set of
coupled self-consistent equations~\cite{Eliashberg,AllenMitrovic,Carbotte,ScalapinoSchriefferWilkins}:
\begin{eqnarray}
&&Z_J(n) =1+\frac{\pi k_BT}{i\omega_n}\sum_{m,J'}\lambda_{JJ'}(n,m)\frac{i\omega_m Z_{J'}(m)}{R_{J'}(m)}\label{el1}\;,\\
&&\Delta_J^{ph}(n) = \pi k_BT\sum_{m,J'}\lambda_{JJ'}(n,m)\frac{\Delta_{J'}(m)Z_{J'}(m)}{R_{J'}(m)}\label{el2}\;,\\
&&\Delta_J^{C}(n) = -\pi k_BT\sum_{m,J'}\mu^*_{JJ'}\frac{\Delta_{J'}(m)Z_{J'}(m)}{R_{J'}(m)}\label{el3}\;,\\
&&R_J(n) = \sqrt{\left(\omega_n^2+\Delta_J^2(n)\right)Z_J^2(n)}\label{el4}\;,\\
&&\lambda_{JJ'}(n,m)=\int d\omega\frac{2\omega\alpha^2F_{JJ'}(\omega)}{(\omega_n-\omega_m)^2+\omega^2}\label{el5}\;.
\end{eqnarray}
Here $\Delta_J(n)=\Delta_J^{ph}(n)+\Delta_J^{C}(n)$ is the total superconducting gap accounting for phononic and Coulombic contributions on the $J$-th FS sheet at frequency $i\omega_n$. 
$Z_J(n)$ is the (phononic) mass renormalization function. This term enters in the diagonal part of the electronic self-energy and contributes both to the superconducting state and the normal state. 
Due to the assumption that all the diagonal contributions stemming from $W$ are already accounted at the level of the normal state Kohn-Sham system, $Z_J(n)$ has a purely phononic character.
 
The off-diagonal Coulombic contributions are accounted by the FS-dependent $\mu^*_{JJ'}$. In this work,  we choose $\mu^*_{JJ'}$ in such a way to reproduce the gap structure obtained within 
SCDFT~\cite{CaC6-nostro,CaC6-H} (we shall come back to this point in Sec.~\ref{sec:res-ELvsSCDFT}).  
Since $\Delta^{ph}( n)$ goes to zero for frequencies much larger than the phononic scale, the cut off $\omega_c$ in the Matsubara frequency introduced for Coulombic terms can be uniformly applied to all the terms of the Eliashberg equations.

$\alpha^2F_{JJ'}(\omega)$ in Eq.~(\ref{el5}) is the band-resolved Eliashberg function.
This quantity results from the averaging of the electron-phonon interaction over FS sheets:
\begin{equation}\label{eq:a2f}
\alpha^2F_{JJ'}(\omega)=\frac{1}{N_J}\sum_{\substack{
 {\bf k}\in J \\
 {\bf k}'\in J'\\
 \nu  }} \delta(\xi_{\bf k})\delta(\xi_{{\bf k}'})\delta(\omega-\omega_{{\bf q},\nu})\left|g^{\nu}_{{\bf k},{\bf k}'}\right|^2\;,
\end{equation}
where $N_J$ is the density of states per spin at the $J$-th FS portion.

Once the gap function $\Delta_J(n)$ and the mass renormalization function $Z_J(n)$ have been obtained on the imaginary axis solving the Eliashberg equations (Eqs.~(\ref{el1})--(\ref{el5})), 
they can be efficiently continued to the real axis via Pad\'e approximant technique~\cite{PadeEliashberg,Baker,Bender}, allowing the computation of the real-axis retarded Green's function. In particular, we will deal with $G := \bar G^{11}$ which is given by:
\begin{equation}
G_J({\bf k},\omega)=\frac{\omega Z_J(\omega)+\xi_{J{\bf k}}}{\left[\omega Z_J(\omega)\right]^2 - \xi^2_{J{\bf k}} -\left[\Delta_J(\omega) Z_J(\omega)\right]^2 },
\end{equation} 
with the corresponding spectral function defined as:
\begin{equation}
A_J({\bf k},\omega)=-\frac{1}{\pi}{\rm Im} G_J({\bf k},\omega).
\end{equation}

The energy and life time of a quasi-particle are given by the real part and (the absolute value of the) imaginary part of the pole positions, respectively. The spectral function may have broader structures and, thus, the concept of quasi-particle may not apply rigorously. 
In order to extract the main features, we look for the zero, $z_p$, of the denominator of $G$, i.e. we look for solutions of the equation:
\begin{equation}
z_p = \frac{\sqrt{\xi^2_{J{\bf k}}+\left[\Delta_J(z_p) Z_J(z_p)\right]^2}}{Z_J(z_p)} \label{eq:qp}
\end{equation}
in the complex plane. 
%========================
\section {COMPUTATIONAL DETAILS}
\label{sec3}
%========================

Electronic eigenvalues, phononinc frequencies, and electron-phonon matrix elements are calculated using the ESPRESSO pseudopotential based package~\cite{pwscf,ESPRESSOreview}. 
All calculations are done using the GGA (Generalized-Gradient Approximation) with the Perdew-Wang~\cite{pw91} parameterization for the exchange-correlation functional. 
Ultrasoft pseudopotentials~\cite{Vanderbilt} are employed.
A $30$ Ry cut-off is fixed for the planewave expansion of the wavefunctions and $300$ Ry for the electronic charge. 
The Brillouin zone is sampled with a $6\times6\times6$ ${\bf k}$-point grid, and electron-phonon matrix elements are obtained on a $10\times10\times10$ grid. More details can be found in Ref.~\onlinecite{CaC6-nostro}. 

The double Brillouin zone integration appearing in the definition of the band-resolved Eliashberg functions in Eq.~(\ref{eq:a2f}) is evaluated with a Metropolis integration scheme. 
Random ${\bf k}$-points are generated on the Brillouin zone, then each ${\bf k}$-point is accepted or rejected with a probability depending on $\xi_{\bf k}$, and its weight is set inversely proportional to the acceptance probability. 
We use a set of about $2\cdot 10^4$ accepted {\bf k}-points per band. 
Then electron-phonon matrix elements on this random mesh are obtained via interpolation from those calculated on the regular grid. 

Eliashberg equations are solved using a $2500~\mbox{meV}$ cut-off of the Matsubara frequencies, with the cut-off $\omega_c$ for the Coulombic interaction equal to $500$ meV. 
Both parameters are much larger than the maximum phonon frequency of CaC$_6$, which is about $200~\mbox{meV}$. 
The number of Matsubara frequencies at each temperature is fixed by the energy cut-offs, 
and the $M$ Matsubara frequencies on the positive imaginary axis are used to construct $(M,M)$ Pad\'e approximants which are used for the analytic continuations\cite{Padecomment} to the real axis.

%========================
\section {RESULTS and DISCUSSION}
\label{sec4}
%========================
\subsection{Properties of the electron-phonon coupling}\label{sec:res-elph}
We calculate the Eliashberg function defined by Eq.~(\ref{eq:a2f}) in three different ways: 
\begin{itemize}
\item Averaging the electron-phonon coupling over the full surface: referred to as $1$-band FS or isotropic approximation.  
\item Splitting the FS into three parts: this is shown with three different colors in the FS (see Table~\ref{tab:lambdaIJ}). 
The division of the Fermi surface leads to a division in the electronic bands shown in the band structure plot of Fig.~\ref{fig:bands};
an electronic band and a portion of the FS have same color if they intersect.
The first portion FS 1 is the external FS sheet (shown in green), which comes from $\pi$ states. 
FS 2 (shown in blue) is the spherical Ca Fermi surface. FS 3 (shown in red) is the $\pi$-prism, a two-dimensional FS having the shape of a hexagonal prism which crosses the spherical Ca Fermi surface (the corresponding  electronic band is the band 3). This division is referred to as the $3$-band FS approximation. 
\item Splitting the FS into six parts: each of the three portions in the $3$-band FS approximation is further split into two parts. 
The above $\pi$ (external) FS has been divided into a less coupled~\cite{CaC6-nostro} outer part with $|${\bf k}$_{xy}|>$ $0.4$ a.u., and the rest; these two portions are referred to as 1a and 1b portion, respectively. 
The Ca spherical Fermi surface is cut into 2a portion (with $|${\bf k}$_{z}|<$ 0.18 a.u.) and the 2b portion (the rest). 3a and 3b
portions for the $\pi$-prism are defined in a similar manner. The same boundaries are used to further split the corresponding energy bands. This overall division is called the $6$-band FS approximation.
\end{itemize}

In Table \ref{tab:lambdaIJ} is presented the DOS at the Fermi level, the intra- and the inter-band electron-phonon couplings:
\begin{equation}
\lambda_{JJ'}=2\int{\frac{\alpha^2F_{JJ'}(\omega)}{\omega}}\label{lambda}d\omega\;.
\end{equation}

\begin{figure}
 \vspace{0.2cm}
 \begin{minipage}{0.5\textwidth}
  \includegraphics[clip= ,width=0.8\textwidth]{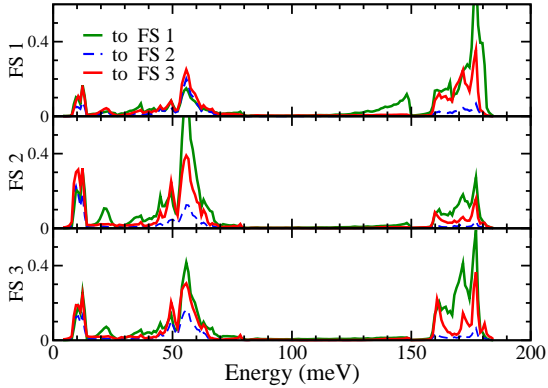}
 \end{minipage}
\caption{ (color online) 3-band FS resolved \aFIJ\ of \cac. The three panels refer to the incoming FS, from which the electron is scattered, while the three lines in each panel refer to the outgoing FS.}\label{fig:a2f}
\end{figure}

In the $3$-band FS approximation the interaction is dominated by the off-diagonal coupling terms, especially by the inter-band scattering from states on the spherical Ca Fermi surface to states on the $\pi$-bands. 
The main reason for this is a strong electron-phonon coupling in the former (from 2 to 1) and large DOS in the latter (from 2 to 3).
The full $3$-band $\alpha^2F_{JJ'}(\omega)$ matrix (see Fig.~\ref{fig:a2f}) shows the distribution of the coupling among the various phonon modes-- band $2$ couples strongly with Ca modes (giving a low frequency peak around $10$ meV) while, band $1$ couples mainly with the high-frequency stretching C modes (at $170~\mbox{meV}$). Band 3 shows the most homogeneous coupling (its intra-band spectral function looks similar in shape to the total Eliashberg function). 

 The further decomposition into the $6$-band FS approximation doesn't introduce qualitative differences with respect to the $3$-band decomposition but, as we shall show, it results in a better quantitative description of the anisotropy of the superconducting properties. 

\begin{table}
\begin{center}
\begin{minipage}{0.5\textwidth}
  \begin{minipage}{0.5\textwidth}
  \begin{flushright}
  \begin{tabular}{c|ccc}
   &   1   &   2   &   3   \\
\hline
1   & 0.301 & 0.136 & 0.257 \\
2   & 0.546 & 0.239 & 0.479 \\
3   & 0.427 & 0.198 & 0.367 \\
\hline
DOS & 0.412 & 0.104 & 0.249 \\ 
  \end{tabular}
  \end{flushright}
  \end{minipage}  %
  \begin{minipage}{0.45\textwidth}
  \begin{center}
     \includegraphics[clip= ,width=0.9\textwidth]{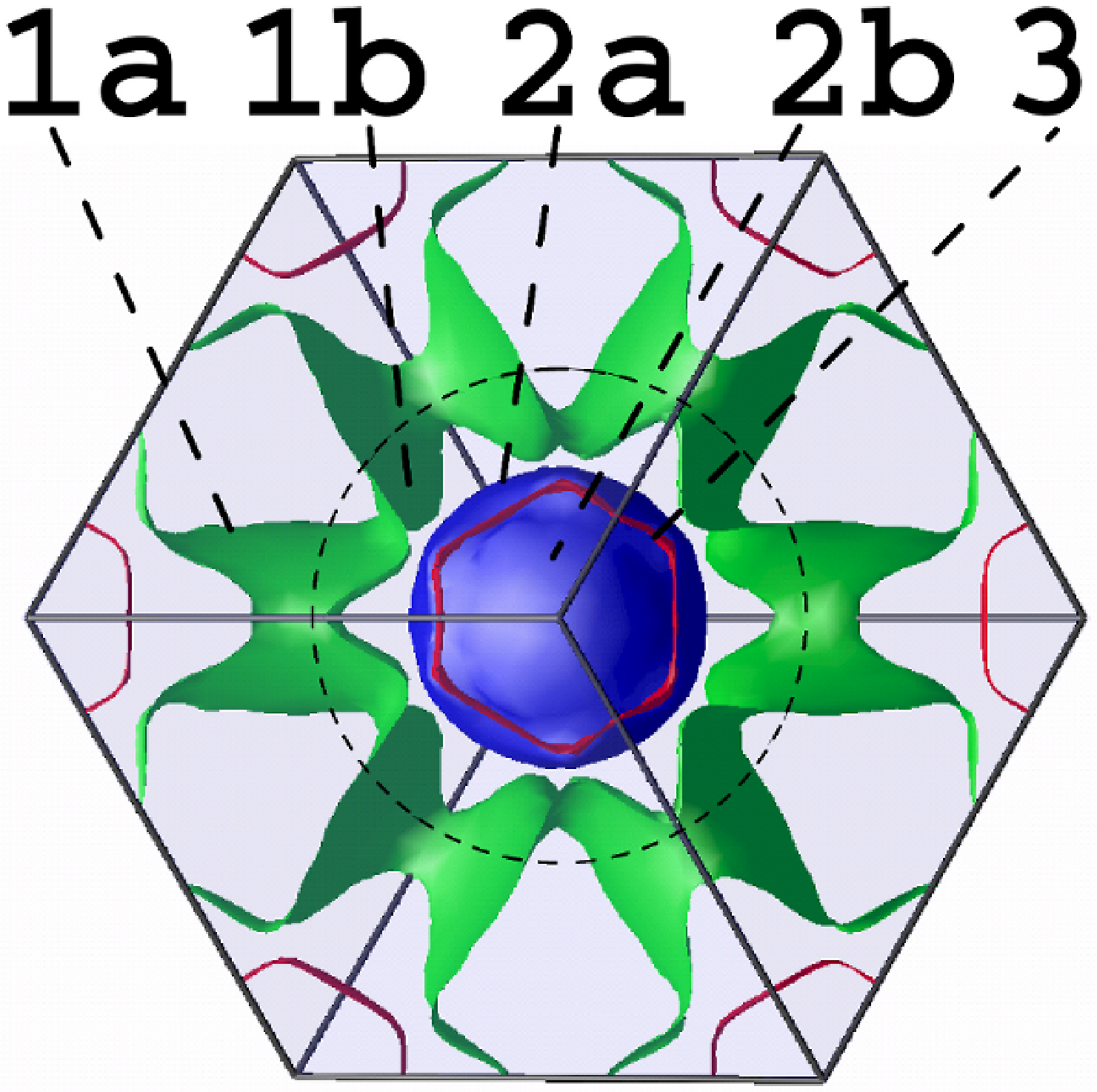}    
  \end{center}
  \end{minipage}
\end{minipage}

\begin{tabular}{c|cccccc}
    &   1a  &   1b  &   2a  &   2b  &   3a  &   3b  \\
\hline
1a   & 0.163 & 0.126 & 0.099 & 0.033 & 0.201 & 0.046\\
1b   & 0.179 & 0.140 & 0.105 & 0.035 & 0.221 & 0.050\\
2a   & 0.331 & 0.245 & 0.151 & 0.084 & 0.384 & 0.096\\
2b   & 0.271 & 0.202 & 0.206 & 0.047 & 0.400 & 0.080\\
3a   & 0.252 & 0.194 & 0.145 & 0.061 & 0.309 & 0.073\\
3b   & 0.206 & 0.157 & 0.128 & 0.044 & 0.259 & 0.060\\
\hline
DOS  & 0.250 & 0.176 & 0.075 & 0.031 & 0.200 & 0.056\\
\end{tabular}
\caption{ Electron-phonon coupling $\lambda_{JJ'}$ and DOS for the $3$-band FS (top) and $6$-band FS (bottom) divisions. 
DOS is given in states/eV/spin.The first index runs over the columns. Isotropic DOS and $\lambda$ are $0.787$ and $0.870$, respectively. 
The picture on the top represents the FS of CaC$_6$: the external  portion in green; the internal sphere  in blue; and the hexagon that cuts the sphere in red. 
The labels ``a" and ``b" correspond to the additional splitting defining the  $6$-band division.}
\label{tab:lambdaIJ}
\end{center}
\end{table}

\subsection{Solution of Eliashberg equations}\label{sec:res-ELvsSCDFT}

\begin{figure}
  \begin{minipage}{0.5\textwidth}
    \includegraphics[clip= ,width=0.8\textwidth]{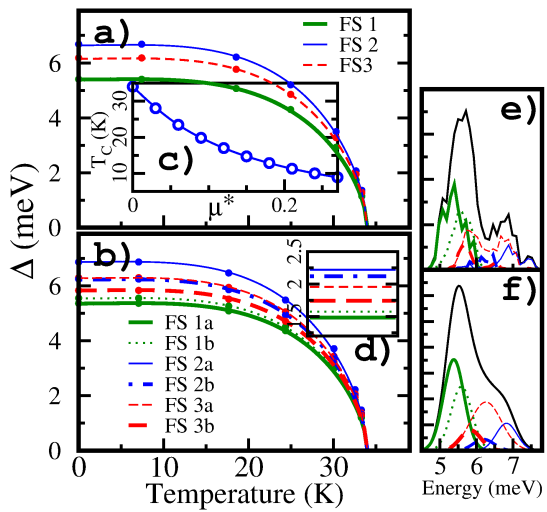}
\caption{(color online) 
(a) Anisotropy of the superconducting gap, in the 3-band FS approximation; and (b) 6-band FS approximation. 
In (a), the dark green thick line is used for the gap computed over FS 1; the blue thin line for  the gap computed over FS 2; and the red dashed line for  gap computed over FS 3. 
In (b), dark green thick and dotted thin lines are used for 1a and 1b FS, respectively; blue thin and dash-dotted thick lines for 2a and 2b FS, respectively; and red  dashed thick and dashed thin lines for 3a and 3b FS, respectively. 
(a) and (b) refer to phonon only calculations
Inset (c) shows the effect of  the inclusion of the Coulombic interaction: the Eliashberg $T_c$ is given as a function of $\mu^*$ (see text for details). 
Inset (d) reports the low-temperature multi-gap structure for $\mu^*=0.21$. 
Panels (e) and (f) report the gap distribution functions (phonon only calculations) in the 6-band FS approximation in SCDFT~\cite{CaC6-nostro} and Eliashberg theory (with a gaussian broadening applied), respectively.} 
\label{fig:gap} 
  \end{minipage}
\end{figure}

The Eliashberg equations are solved in three different ways corresponding to three ways in which the Eliashberg function defined in section \ref{sec4}.
The solution to the Eliashberg equations lead to a strong anisotropy in the gap (see Fig.~\ref{fig:gap}); the smallest gap corresponds to the external FS (FS 1), while the highest value of the gap is related to the 2a structure which forms the central part of the Ca spherical Fermi surface. We note that at the phononic level, the anisotropic structure obtained here agrees very well with the one obtained within SCDFT in Ref.~\onlinecite{CaC6-nostro} (see panels (e) and (f) in Fig.~\ref{fig:gap}).
If the $T_c$ is determined from this gap function, without including the Coulomb interaction, it is not strongly affected by the multi-band character (i.e. the anisotropy of the gap function); the isotropic $T_c$ is about $33.5~\mbox{K}$ and only slightly higher in the 6-band case with a value of $34~\mbox{K}$.

In order to include the Coulomb interactions the matrix $\mu^*_{JJ'}$ is needed. 
Typically $\mu^*_{JJ'}$ is determined by fitting to the experimental data. However, for \cac\ the SCDFT gap well   reproduces the experimental measurements \cite{DagheroGonnelli_pcs,CaC6-Gonnelli,Nagel,Shiroka,DagheroGonnelliRev,Kurter}. Therefore, in this work, we determine $\mu^*_{JJ'}$ fitting to the gap structure obtained from SCDFT.   

Interestingly, the simple \textit{semi-isotropic} approximation: $\mu^*_{JJ'}=\mu^* N_{J}/N$, turns out to be sufficient to reproduce the SCDFT gap.
In the $6$-band FS approximation the experimental $T_c = 11.5~\mbox{K}$ is reproduced for $\mu^* \simeq 0.21$ [see inset (c) in Fig.~\ref{fig:gap}]. 
With this choice of $\mu^*_{JJ'}$ the inclusion of Coulombic effects reduces the $T_c$ without significantly affecting the anisotropic structure of the superconducting gap. 
The only difference is that by including the Coulombic interaction the gap corresponding to the 2b portion of the FS becomes slightly larger than the 3a portion (see Fig.~\ref{fig:gap}(d)). 
This choice of a semi-isotropic Coulombic pairing which is often done in Eliashberg theory can be validated by the present analysis. 
However, this cannot be applied as a general rule; it has been shown, in cases like in MgB$_2$~\cite{MgB2-PRL,MgB2-review}, that a more detailed Coulombic structure is necessary to get the correct gap anisotropy. 

\begin{figure}
 \vspace{0.2cm}
 \begin{minipage}{0.5\textwidth}
  \includegraphics[clip= ,width=0.8\textwidth]{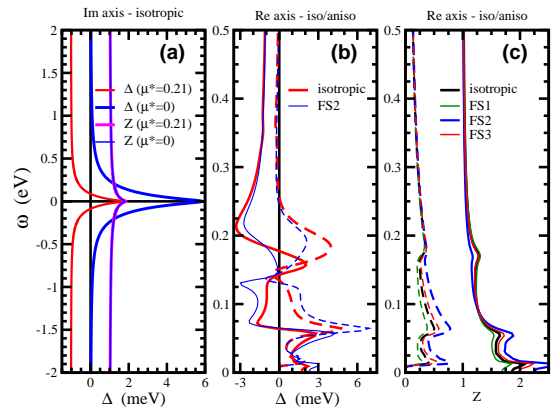}
 \end{minipage}
\caption{(color online) Eliashberg functions. 
 (a) Superconducting gap and mass renormalization function calculated solving the isotropic Eliashberg equations on the imaginary axis 
(b) Analytically continued gap function on the real axis. (c) Analytically continued mass renormalization function on the real axis. Full lines are used for  the real part and dashed lines for the imaginary part of these complex functions. On the imaginary axis the functions are purely real.}  
\label{fig:eliashberg_functions}
\end{figure}

Both $\Delta(n)$ and $Z(n)$ are purely real-valued on the Matsubara frequencies. For the isotropic case the frequency dependence of $\Delta(n)$ and $Z(n)$ 
 is shown in Fig.~\ref{fig:eliashberg_functions}(a). It is clear that  $Z(n)$ has a value of $1+\lambda$ for small $|n|$ and then monotonically decreases to 1 at energies much larger than the available phonons. This behavior is almost independent of the values of $\mu^*$ and temperature. The $\Delta(n)$ function also monotonically decreases as a function of increasing energy. The low-energy value is the fundamental superconducting gap, while in the high-energy limit approaches $\Delta^C(n)$. For an isotropic and static Coulomb interaction, $\Delta^C(n)$ is both independent of $\bf k$ and the Matsubara frequencies. 
More physical features emerge from the analytic continuation to the real axis. In particular, we see a three-peak structure in both the real and imaginary parts of $Z(n)$ that correlates with the peaks in the \aFIJ.

\subsection{Analysis of self-energy effects}\label{sec:res-SEeffects}

\subsubsection{Normal state: isotropic approximation}\label{sec:pol1}

We discuss in this section the simplest case in which the Eliashberg equations are solved above $T_c$.
Since $T_c$ is quite low with respect to the phonon energies, solution above $T_c$ is almost equivalent to imposing the solution at $\Delta=0$ when $T=0$.
In this section, we use only the isotropic solutions of the Eliashberg equations and a parabolic band dispersion.

\begin{figure*}
  \begin{minipage}{0.7\textwidth}
    \includegraphics[clip= ,width=\textwidth]{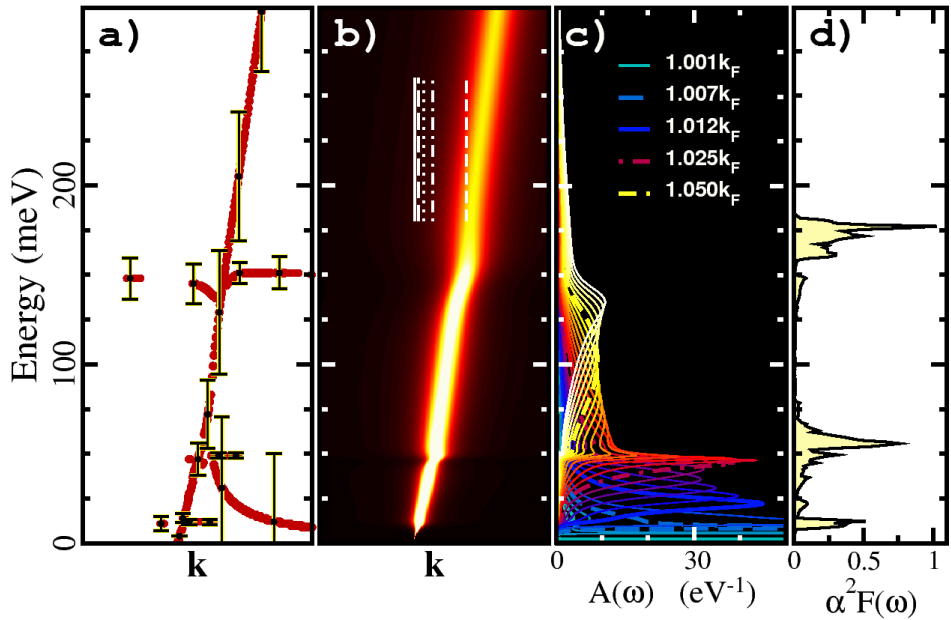}
  \end{minipage}
  \begin{minipage}{0.9\textwidth}
\caption{(color online) Polaronic bands and spectral function for the normal state in the isotropic approximation. On panel (a), the polaronic quasi-particle dispersion curves are shown. 
This is obtained from the real and imaginary part of the solution of  Eq.~(\ref{eq:qp}).   The inverse lifetime of the state
is  given by $|Im(z_p)|$ and it is reported for some points as ``error-bar". 
Panel (b) shows the spectral function $A(k,\omega)$. The abscissa represent the $|{\bf k}|$ axis, and the point in which $A(k,\omega)$ crosses is ${\bf k}_F$. In ordinate, the frequency axis is reported. 
The color-scale goes from zero to about $15$ eV$^{-1}$ (note that, we cut it off  to enhance the structures of $A(k,\omega)$). 
Panel (c) shows $A(k,\omega)$ for different values of the momentum. A few values are highlighted and correspond to $\bf k$-values indicated in panel (b).
Panel (d) shows  \aF\  computed in the the isotropic 1-band approximation. 
}
\label{fig:oneband}
  \end{minipage}
\end{figure*}

The solution of the Eliashberg equations give $Z(\omega)$ which is shown in Fig.~\ref{fig:eliashberg_functions}(c)  (black line). Using  $Z(\omega)$ as input, Eq.  (\ref{eq:qp}) is solved in order to obtain the quasi-particle dispersion curves reported in Fig.~\ref{fig:oneband}(a). 
This figure shows how the unperturbed electronic band, by getting dressed with the phononic self energy, develop branches in correspondence with the three main phononic peaks in the Eliashberg function.
These electronic quasi-particles dressed by the electron-phonon interaction, so called polarons, are nearly dispersionless. Only one dispersive branch, that goes to zero from about $50$ meV, is observed. This structure also has a very short lifetime ($\sim 100$ meV). All the other polaronic modes have instead a longer lifetime (between $1$ and $10~\mbox{meV}$) and thus appear as sharp quasi-particles.
However, the polaronic branches, carry very little of the total spectral weight. Most of the spectral weight is still localized near the bare electron dispersion line. This can also be seen in Fig.~\ref{fig:oneband}(b), where the spectral function is shown in the same energy/momentum window as the quasi-particle plot. In this case, we see how the main electron band acquires a finite lifetime and instead of branchings only kinks appear. These kinks correspond to the three main peaks in the \aF. 

To move from a qualitative description to a more quantitative one, the spectral function $A(\omega)$ is examined (see Fig.~\ref{fig:oneband}(c)). 
At ${\bf k}={\bf k}_{\rm F}$ more than $50$\% of spectral weight is accounted for by a single peak of infinite lifetime at the Fermi energy.
This peak (single green line in Fig.~\ref{fig:oneband}(c)) moves to higher energies with increasing ${\bf k}$-vector (slowly growing in width) up to an energy of about $10~\mbox{meV}$, where it merges 
with the polaronic branch generated by the low-frequency Ca modes (blue long-dashed line). 
Above $10~\mbox{meV}$ the peak is broader (blue thick line) because the electrons can relax through the generation of Ca phonons. 
This broad peak then behaves in a similar way as the narrow peak below $10$ meV; i.e., it increases in energy with ${\bf k}$ up until it merges with another polaronic band which originates from the low-frequency C modes (red dot-dashed line) and has an energy $\sim 50$ meV. 
It becomes very broad (yellow short-dashed line) and is difficult to follow as it merges with the high-frequency C mode. 
This behavior is very similar to the case of the Einstein phonons discussed by Engelsberg and Schrieffer in Ref.~\onlinecite{EngelsbergSchrieffer}. In our case, this is due to the three-peak structure of the \aF\ i.e. due to the combined effect of the two-dimensionality of graphite along with the presence of weakly bound Ca ions.
Therefore, at the isotropic level the self-energy effects in \cac\ (type described by Eq.~(\ref{eq:sigma})~) are particularly simple.

\begin{figure*}[H]
\begin{minipage}{0.99\textwidth}
      \includegraphics[clip= ,width=\textwidth]{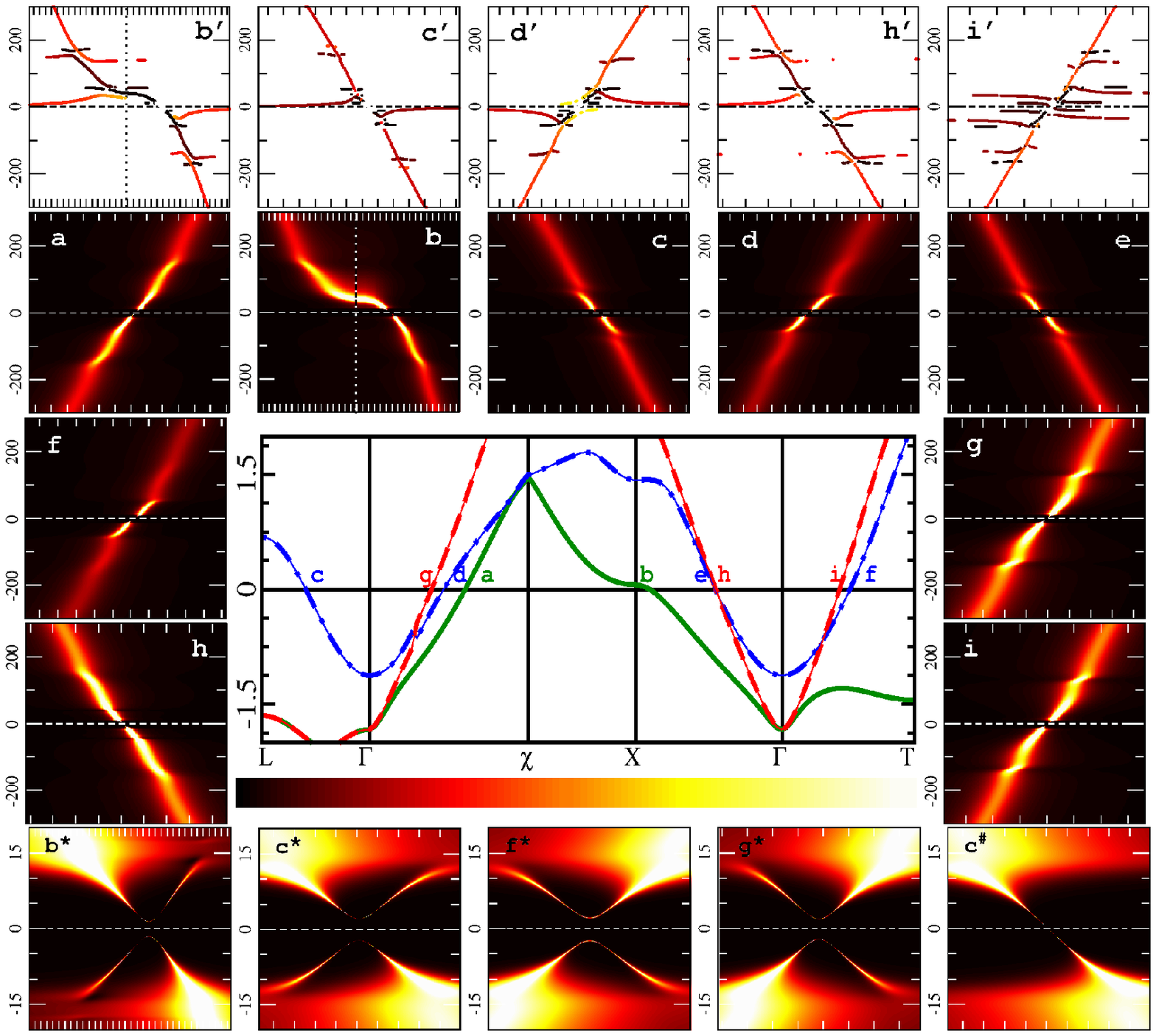}
  \end{minipage}
\begin{minipage}{0.9\textwidth}
\caption{(color online)  
Polaronic bands and spectral function for the anisotropic and superconducting states of CaC$_6$ determined within the 6-band approximation.
Central panel shows the band structure of \cac: the continuous green line is the C$\pi$ band corresponding to FS 1; 
the blue dot-dashed line is the Ca inter-layer band corresponding to FS 2; and the red dashed line is the second C$\pi$ band corresponding to FS 3
(see Table~\ref{tab:lambdaIJ} for plots of the several FS portions). Further subdivisions are not reported explicitly.
Panels from  (a) to (i) show the spectral function near the Fermi energy in a wide energy window of  $300~\mbox{meV}$. 
The letters correspond to the labeling of the crossing points in the central panel. The color-scale goes from zero 
(black) to $15$ eV$^{-1}$ (white). Higher peaks in the spectral function have been cut off.  
Panels with a primed index show the quasi-particle spectrum (see text for details). These also show the polaronic branches together with the main electronic band.  
Colors are proportional to the line-width of the quasi-particle state: black corresponds to zero and yellow/white correspond to about $0.1~\mbox{eV}$. 
Panels (b$^*$), (c$^*$), (f$^*$), and (h$^*$) -- focusing on the superconducting gap (a logarithmic color-scale is used) -- 
show a zoom of the spectral function near the Fermi energy in a narrow energy window of $20~\mbox{meV}$. 
Panel  (c$^\#$) is the same as (c$^*$) but evaluated in the non-superconducting state.
}
\label{fig:bands}
\end{minipage}
\end{figure*}

\subsubsection{Normal and superconducting states: anisotropic features}\label{sec:pol2}
 
The degree of complexity of our analysis is increased by making use of the real KS band dispersions of CaC$_6$~\cite{CaC6-nostro,CalandraMauri}. 
Multiple FSs that couple with different phonon branches are accounted for by adopting the 6-band decomposition.
As shown in Fig.~\ref{fig:bands}(c), (d), (e), and (f), the Ca band couples mostly with low-frequency modes, therefore it shows polaronic structures only up to $50~\mbox{meV}$.  One single kink can be observed in the spectral function at $50~\mbox{meV}$. The kink at 10 meV is not visible simply because below this frequency the spectral function itself is just a sharp peak.

The band that has most structures is the one that produces the $\pi$-prism FS (Fig.~\ref{fig:bands}(g), (h), and (i)), due to the coupling of  all three sets of modes. 
The external FS, which couples mostly with high-frequency C modes,  shows only a weak kink around $160~\mbox{meV}$. 
The polaronic branchings (Fig.~\ref{fig:bands}(b'), (c'), (d'), (h'), and (i')) have similar structures as observed in the isotropic limit. 
Anisotropic features are less marked than in the spectral function, because Eq.~(\ref{eq:qp}) used  to determine these features does not retain information about the spectral weight in the branches. 

The spectral function in the superconducting phase is gapped and the multi-gap features can be clearly seen in the lower panels of Fig.~\ref{fig:bands}(b$^*$), (c$^*$), (f$^*$), and (g$^*$). The gap ranges from $1.5~\mbox{meV}$ around the point b in the band structure (Fig.~\ref{fig:bands}(b$^*$)) to about $2.2~\mbox{meV}$ 
around point c (Fig.~\ref{fig:bands}(c$^*$)). 
The spectral function shows a (textbook-like) hyperbolic dispersion-- this is the signature of Bogoljubov excitations\cite{Schrieffer}, as compared to the normal electronic excitations which have a linear band dispersion (see Fig.~\ref{fig:bands}(c$^*$) and (c$^\#$)).
The distance between the vertices of the hyperbola is equal to $2\Delta(0)$. As the temperature rises to approach $T_c$,  $\Delta(0)\to0$ and the hyperbola tends to its asymptotes.
At the same time the spectral weight of the two reflected components (right part of the upper branch and left  part of the lower branch  in Fig.~\ref{fig:bands}(c$^*$)) also goes to zero and the excitation spectrum becomes normal (Fig.~\ref{fig:bands}(c$^\#$)).
 
One of the two arms of this hyperbolic dispersion corresponds to the normal electronic dispersion line, and it behaves in a similar way as in the non superconducting state. The other arm that is a unique feature of the Bogoljubov excitations loses spectral weight as the distance from the FS increases.
However as it reaches the energy of the Ca in-plane modes (from $10$ to $15~\mbox{meV}$) it deviates from the hyperbolic arm and follows a polaronic (dispersionless) behavior. One can appreciate the formation of this dispersionless Bogoljubov polaron in the upper-right and lower-left corners of panel (b$^*$) in Fig.~\ref{fig:bands}.

%========================
\section {CONCLUSIONS}
\label{sec5}
%========================
Superconductivity  in the graphite intercalated compound \cac\  is studied using Eliashberg theory and superconducting  density functional theory.  
Within a multi-band description and assuming a structureless Coulomb interaction, we performed a detailed analysis of the influence of strongly anisotropic electron-phonon coupling on the {\bf k}-dependence of the superconducting gap. 
Anisotropies computed with Eliashberg theory and superconducting density functional theory were found to be in very good agreement with each other and with experiments~\cite{CaC6-Gonnelli} .

In this context, from  the solution of the Eliashberg equations, we have shown how anisotropic polaronic bands emerge over different Fermi surface sheets. 
The  interplay between superconducting (Bogoljubov) excitations and polarons has also been studied. 
We reported, for the first time, how Engelsberg-Schrieffer polarons  evolve from the normal state to the superconducting state in CaC$_6$.  

%========================
\section{Acknowledgments}
A.S. acknowledges useful discussions with A. Eiguren. 
S.P. acknowledges support through DOE grant DE-FG02-05ER46203. 
S.M. acknowledges support by the Italian MIUR through Grant No. PRIN2008XWLWF9.
\label{ack}
%========================


\begin{thebibliography}{99}
\bibitem{Padecomment}
Pad\'e approximants were used to continue solutions of the Eliashberg equations to the real axis. 
Very good agreement is observed in the full energy spectrum for the $Z(\omega)$ function.
Although the analytically continued gap function $\Delta(\omega)$ shows some deviations from the exact solutions, this happens in the $\sim0.1~\mbox{eV}$ energy region\cite{PadeEliashberg}. As a result, this doesn't introduce  sizeable errors in our estimation of the spectal function.
\end{thebibliography}
\end{document}